\documentclass[]{aastex631}

\usepackage{graphicx}    
\usepackage{amsmath}
\usepackage{float}

\usepackage{placeins}
\shorttitle{SNR 0509-67.5 progenitor system}
\shortauthors{Das et al.}

\begin{document}

\title{Deep MUSE observations of SNR 0509-67.5 reveal a double degenerate merger progenitor}

\author[0000-0002-5483-0232]{Priyam Das}
\altaffiliation{E-mail: priyam.das@unsw.edu.au}
\affiliation{School of Science, The University of New South Wales, Northcott Dr., Canberra, ACT 2612, Australia}
\affiliation{Research School of Astronomy and Astrophysics, Australian National University, Canberra ACT 2611, Australia}

\author{Ivo R. Seitenzahl}
\affiliation{Research School of Astronomy and Astrophysics, Australian National University, Canberra ACT 2611, Australia}
\affiliation{Mathematical Sciences Institute, Australian National University, Canberra ACT 2611, Australia}

\author{Gilles Ferrand}
\affiliation{The University of Manitoba, Department of Physics and Astronomy, Winnipeg, Manitoba, R3T 2N2, Canada}
\affiliation{RIKEN Center for Interdisciplinary Theoretical and Mathematical Sciences (iTHEMS), Wak\={o}, Saitama 351-0198 Japan}

\author{R\"udiger Pakmor}
\affiliation{Max-Planck-Institut f\"ur Astrophysik, Karl-Schwarzschild-Str. 1, 85748 Garching, Germany}

\author{Simon J. Murphy}
\affiliation{School of Science, The University of New South Wales, Northcott Dr., Canberra, ACT 2612, Australia}

\author{Ashley Ruiter}
\affiliation{Mathematical Sciences Institute, Australian National University, Canberra ACT 2611, Australia}
\author{Brian P.~Schmidt}
\affiliation{Research School of Astronomy and Astrophysics, Australian National University, Canberra ACT 2611, Australia}
\author{Friedrich K.~R\"opke}
\affiliation{Heidelberger Institut für Theoretische Studien, Schloss-Wolfsbrunnenweg 35, 69118 Heidelberg, Germany}
\affiliation{Zentrum für Astronomie der Universität Heidelberg, Institut für Theoretische Astrophysik, Philosophenweg 12, 69120 Heidelberg, Germany}
\affiliation{Zentrum für Astronomie der Universität Heidelberg, Astronomisches Recheninstitut, M{\"o}nchhofstra{\ss}e 12--14, 69120 Heidelberg, Germany}

\begin{abstract}
Deep MUSE observations of SNR 0509-67.5 reveal that the coronal [Fe\,\textsc{xiv}] $\mathrm{\lambda}$5303 emission line appears with either one or two velocity components across the entire remnant, arising from reverse-shocked ejecta moving toward and away from the observer. A supervised dense neural network classifies each spaxel and fits Gaussian profiles plus a linear function to the observed line emission. We measure a bulk Doppler velocity of $-1000\pm60~~\mathrm{km~s^{-1}}$, interpreted as the line-of-sight component of the primary white dwarf's orbital velocity in a double-degenerate merger. The red- and blue-shifted ejecta map shows a flattened  edge along the north-eastern rim, consistent with the companion's shadow, indicating a binary companion was present at explosion. Modelling this feature as a cone anchored at the explosion centre and applying Bayesian inference, we recover the cone's orientation and half-opening angle. We then use the Eggleton Roche-lobe relation to infer properties of the companion. The companion was likely a ${\sim}0.6~\mathrm{M_\odot}$ white dwarf with radius ${\sim}9800$~km and orbital velocity ${\sim}1700~\mathrm{km~s^{-1}}$ at the time of explosion. Together, these results provide a complete dynamical picture of a Type Ia supernova progenitor system whose maximum-light spectrum is independently constrained by light echo observations.
\end{abstract}

\keywords{ISM: supernova remnants -- machine learning -- white dwarfs -- binaries -- supernovae}

\section{Introduction}
Type Ia supernovae (SNe Ia) are explosive deaths of carbon-oxygen white dwarfs (WDs). They act as standardizable candles \citep{Phillip-1993a}, which are used to study the accelerated expansion of the Universe \citep{Schmidt1998, Perlmutter1999a}. They also produce more than half of the iron-peak elements in the Universe and some intermediate mass elements. Despite their cosmological importance \citep{Branch1998, Riess1998, lu2022}, the origin of SNe Ia and their progenitor systems are not yet fully understood and the topic remains one of the long standing mysteries in stellar evolution \citep{Mazzali2007}. The general consensus is that the progenitors are binary (or possibly triple+) star systems before the cataclysmic explosion. However, the progenitor systems of SNe Ia remain highly debated, with multiple binary scenarios having been proposed, most notably the single-degenerate and double-degenerate channels. In the single-degenerate scenario, the primary WD shares a binary orbit with a non-degenerate He-burning or H-burning companion star. In the double degenerate scenario, the companion is another WD \citep{webbink84, Kerkwijk2010}. The two WDs either collide head on, or (more likely), the explosion is triggered in a merger scenario \citep[see][for more details]{Ruiter2025}. However, due to the non-detection of a clear progenitor system or surviving companion, direct observational constraints on the exact progenitor scenario remain elusive \citep{Kerzendorf2009,Schaefer2012, Pan2014,Shields2023}.
Type Ia supernova remnants, however, provide a unique opportunity to constrain the progenitor system at later stages, when the ejecta have expanded and evolved into a supernova remnant.\\
Type Ia supernova remnants (SNR Ia) evolve in a partially neutral and  very low ambient density medium, creating Balmer dominated shock regions due to the interaction of the forward shock with the ambient medium \citep[see][for detailed review]{Heng10}. Most importantly, the interaction of the stellar ejecta with the ambient medium initiates a reverse shock, propagating inwards through the ejecta in the Lagrangian sense. The reverse shock interacts with the ejecta, heating them to high temperatures suitable for emitting X-rays. This process in turn produces emission lines in the X-ray that directly traces the chemical composition, density, temperature and ionization state of the ejecta \citep{badenes2003,badenes2005,albert2022}. Recent work has revealed that the reverse-shocked ejecta also emit in broad coronal lines from highly-ionized atoms in the optical \citep{seitenzahl2019}. With high resolution spectroscopy of young SNR Ia in the optical, the ejecta morphology of these remnants can be studied in great detail. Mapping the ejecta is particularly important because it traces the stellar material directly from the explosion. Some 3D hydrodynamical remnant models have been created from forward modeling the 3D hydrodynamical explosion models to match the observed ejecta properties \citep{Orlando2012,Orlando2016,Ferrand2019}. The simulated 3D remnant models of the delayed-detonation and the double-detonation explosion mechanisms show inherent differences in the geometry and morphology of the ejecta distribution in each case \citep{khokhlov1991a,Roepke2007a,Roepke2007b,maeda2010nucleosynthesis,seitenzahl2013a,Tanikawa2019,pakmor2022a}. These differences still persist in the SNR phase, allowing the observed ejecta distribution in young remnants to be directly compared with the models and link back to their underlying explosion mechanism. \citep{zhou2018,Uchida2024}. This technique has already revealed clues regarding the explosion mechanism of SNR 0509-67.5 (SNR 0509) \citep{Das2025, Mandal2026,Soker2025a, Das2026}, which stemmed from a 1991T-like luminous SN~Ia as revealed by maximum light spectra observed from the SN’s light echo reflecting off dust along the line of sight \citep{Rest2008}. 

In this Letter, we analyze the global Doppler velocity morphology of the reverse-shocked ejecta in the young Type Ia supernova remnant SNR 0509-67.5 using deep MUSE observations. By reconstructing the three-dimensional ejecta geometry from the Doppler-split [Fe\,\textsc{xiv}] emission, we identify a shadow effect consistent with a double-degenerate merger progenitor system and derive constraints on the orbital properties of the progenitor binary pre-explosion. Section 2 describes the observational data and analysis methodology, while Section 3 presents the results and their implications for the progenitor scenario.

\section{Methods}
\subsection{Data}
In this work, we use deep MUSE IFS observations of SNR 0509–67.5 (${{\sim}}$30 h) obtained under P.ID 0104.D-0104(A) (PI: Seitenzahl). The observations cover a ${{\sim}}1'\times1'$ field of view with a spatial sampling of $0.2''\times0.2''$, spanning a wavelength range of ${{\sim}}4750$–9350 \AA\ at a spectral resolving power of $R{{\sim}}$ 3000. The data acquisition and reduction is presented in \citet{Das2025}. 

\citet{Das2025} observed splitting of the broad coronal emission line from [Fe\,\textsc{xiv}] in the reverse-shocked ejecta (see their Figure~1), which we interpret here as the blue-shifted and red-shifted face of the expanding remnant. The line profile remains predominantly single-peaked near the rim, where the ejecta motion is largely in the plane of the sky, but separates into two distinct components with increasing separation as we move from the rim of the remnant towards the centre. We use a supervised machine learning approach to identify whether the broad [Fe\,\textsc{xiv}] emission line in each spaxel contains one or two velocity components and to identify their approximate peak positions, which are then used as initial conditions for fitting one- or two-component Gaussian profiles with a linear continuum.  We then compute the Doppler velocity of these lines and reconstruct the [Fe\,\textsc{xiv}] ejecta Doppler velocity distribution in 3D. 

\subsection{Artificial neural network architecture}

We use a dense Interconnected Neural Network (INN) \citep{Rhea2023, Baron2024, Bracci2025, Belfiore2025} to analyze each spaxel across the full extent of the MUSE cube (${{\sim}}300 \times 300$) and determine whether the spectrum within the wavelength range ($\lambda5200$-$\lambda5450$) containing the broad coronal [Fe\,\textsc{xiv}] $\lambda5303$ emission line exhibits a single peak, double peaks, or no broad emission feature. The network consists of four layers: an input layer, two interconnected hidden layers, and an output layer. The first hidden layer contains 128 neurons with rectified linear unit (ReLU) activation functions \citep{Nair2010}, followed by a dropout layer with a probability of 0.1 to reduce overfitting \citep{Srivastava2014}. The second hidden layer consists of 64 neurons with ReLU activation and an additional 0.1 dropout layer. The model was implemented in \texttt{python (v3.13.0)} using \texttt{tensorflow (v2.20.0)} \citep{tensorflow2015-whitepaper}, and was trained and tested on synthetic spectra generated using one- and two-component Gaussian line profiles.  Finally, the output from layer 2 is fed to the final output layer by the network. Our output layer consists of two shared heads: a) Classifying the number of peaks as no peak (Class 0), one peak (Class 1), two peaks (Class 2) and narrow peak (Class 3), b) Assigning wavelengths to the peaks for both the single- and double peak scenarios. We designed the classification head with 4 neurons, corresponding to 4 classes with a softmax activation function \citep{sharma2017activation}. The regression head is designed with two neurons with a sigmoid activation function that identifies the normalised amplitudes of up to two peaks. 
The final two-head loss function is the weighted sum of the loss functions from both the heads given by $L_\mathrm{total}=L_\mathrm{cls}+ 0.5L_\mathrm{reg}$.
The model is trained on 100000 synthetic spectra for 50 epochs with 20\% of the data used for validation. We implemented the Adam optimizer algorithm \citep{Kingma2014} to train the network, which was built using \texttt{keras v3.10.0} \citep{Keras2018} from \texttt{tensorflow}. The performance of the network in classifying the number of broad [Fe\,\textsc{xiv}] emission peaks is shown by the confusion matrix in Fig.~S1, while the root-mean-square error (RMSE) in identifying the peak wavelengths is presented in Fig.~S2. 

\subsection{Emission line peak detection and curve fitting}
After training, the network is applied to classify the number of peaks in each spaxel across the entire MUSE cube within a defined wavelength range (where [Fe\,\textsc{xiv}] is expected between 5200~\AA\ and 5450~\AA). The network returns a class prediction along with the estimated peak positions. This output is then passed to the next stage of the pipeline, where the prediction is used to select and fit either a single- or double-Gaussian model. The spectra are then fit using Bayesian inference rather than least-squares optimisation, with the parameter posterior sampled via \texttt{emcee}. The predicted peak positions from the regression output head of the network act as priors.  As a measure of fit quality, we additionally record the mean log-posterior across the accepted samples for each spaxel, which is later mapped spatially to identify regions of high and low fit confidence (see Fig.~S3).

\begin{figure*}
    \centering
    \includegraphics[width=1\linewidth, trim={50 80 50 80}, clip]{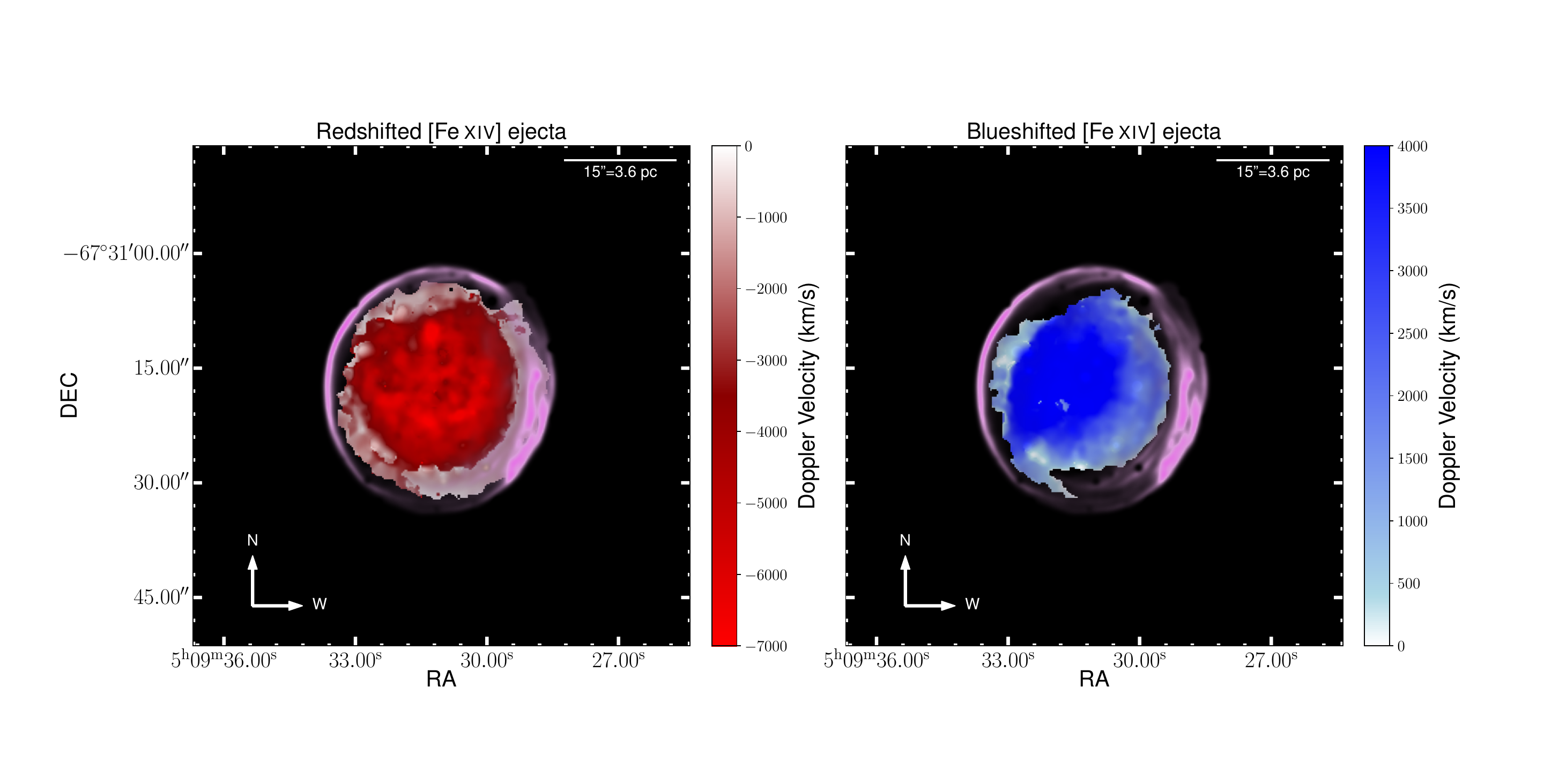}

    \caption{Doppler maps of ejecta emitting in [Fe\,\textsc{xiv}] in SNR 0509, calculated using a rest wavelength of 5308~\AA, are shown separately. {\bf Left:} The red-shifted ejecta, showing lower Doppler velocities along the edges. {\bf Right:}  The blue-shifted component of the ejecta, which also shows lower Doppler velocities in the western region of the remnant, likely linked to a denser ambient medium. H$\alpha$ is overlaid in magenta to show the full extent of SNR 0509.}
    \label{fig:doppler_maps}
\end{figure*}

The output of the full pipeline consists of the fitted parameters, including the Doppler velocity of each individual peak, the bulk Doppler velocity inferred from the midpoint of the two peaks, surface brightness, velocity widths, confidence score, a Doppler map for single-peak spaxels, and a bulk Doppler map for double-peak spaxels derived from the peak midpoints. The bulk Doppler map  traces the line-of-sight motion of the emitting shell as a whole rather than the expansion of its front and back surfaces.
During post-processing, we apply a Kriging interpolation — a geostatistical method that estimates missing values from the spatial covariance of surrounding data — to fill pixels where the pipeline failed to fit a Gaussian profile due to stellar contamination, using the \texttt{pykrige} module \citep{Murphy2014}. We adopt ordinary Kriging, as it utilises a variogram to model spatial autocorrelation, allowing it to capture complex spatial variations in the data rather than applying a simple smoothing algorithm. Following interpolation, the blue-shifted and red-shifted velocity maps are visualised using \texttt{matplotlib} \citep{hunter2007matplotlib} as shown in Fig~\ref{fig:doppler_maps}. The apparent discontinuity between the outer low-velocity envelope and the inner high-velocity core marks the spatial boundary at which the [Fe\,\textsc{xiv}] profile becomes resolvable into distinct approaching and receding components. In the outer region, the line is fit with a single Gaussian tracing the bulk line-of-sight velocity of unresolved double peaks, whereas the inner region exhibits the double peak characteristics.

\subsection{Reconstructing 3D ejecta morphology}
\begin{figure}
    \centering
    \includegraphics[width=0.9\linewidth, trim={450 100 400 200}, clip]{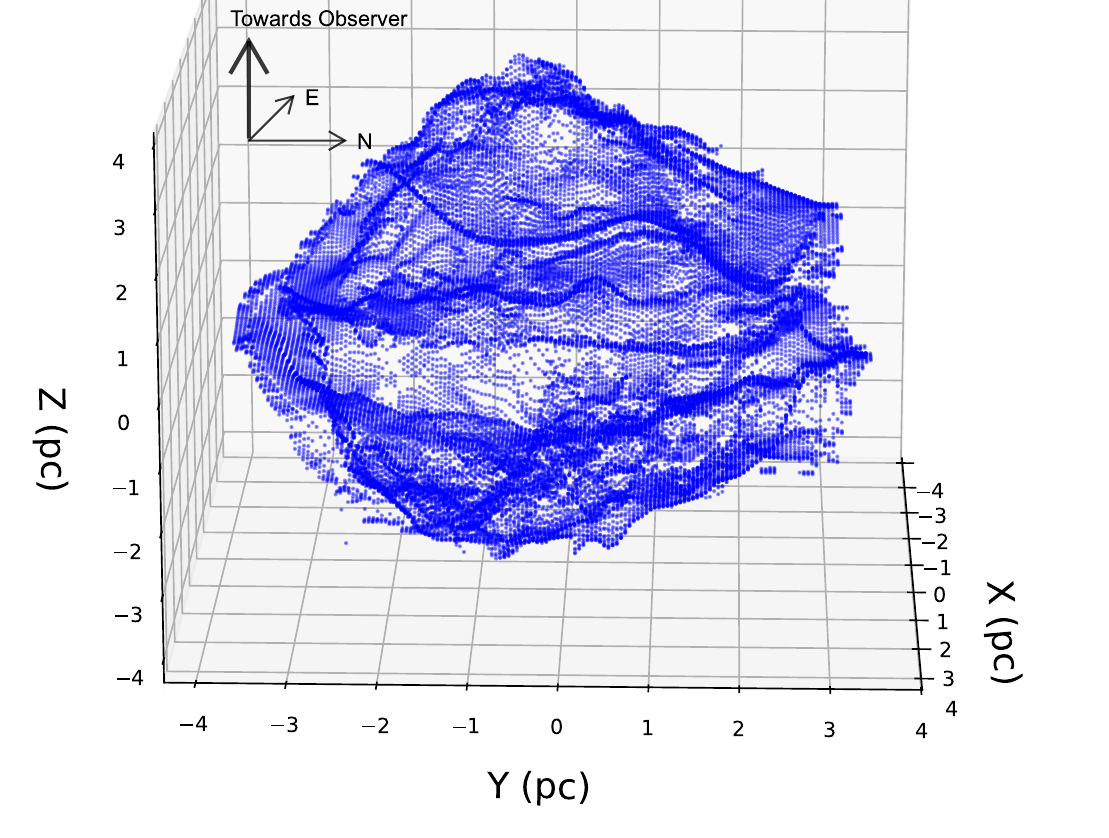}
    \caption{Three-dimensional spatial map of the [Fe\,\textsc{xiv}] ejecta distribution in SNR 0509 reconstructed from the velocity map. The ejecta distribution shown is along the x-axis (RA).}
    \label{fig:3D_space}
\end{figure}

We use the 3D Doppler velocity distribution of [Fe\,\textsc{xiv}] to reconstruct the ejecta geometry in 3D space. We computed a fraction map corresponding to $\cos(\theta)$ from the maximum Doppler velocity, assuming homologous expansion from the dynamical center ($\alpha$=$5^h09^m31^s.16$, $\delta$=$-67^\circ31'17''.1$; \citealt{arunachalam2022a}), using $\cos\theta = V_{\mathrm{obs}}/V_{\mathrm{max}}$. To account for departures from spherical symmetry, we binned the field into 360 one-degree azimuthal sectors and measured the maximum radial extent in each sector. The line-of-sight (LOS) extent is $z=R~\cos(\theta)$ (where $R$ is the maximum shell radius for each sector) and the projected coordinates are scaled by 0.048~pc per pixel (spatial resolution of MUSE for LMC). This yields a 3D geometry of the ejecta distribution for SNR 0509 (see Fig.~\ref{fig:3D_space}).

\begin{figure*}
    \centering
    \includegraphics[width=1\linewidth, trim={50 10 0 0}, clip]{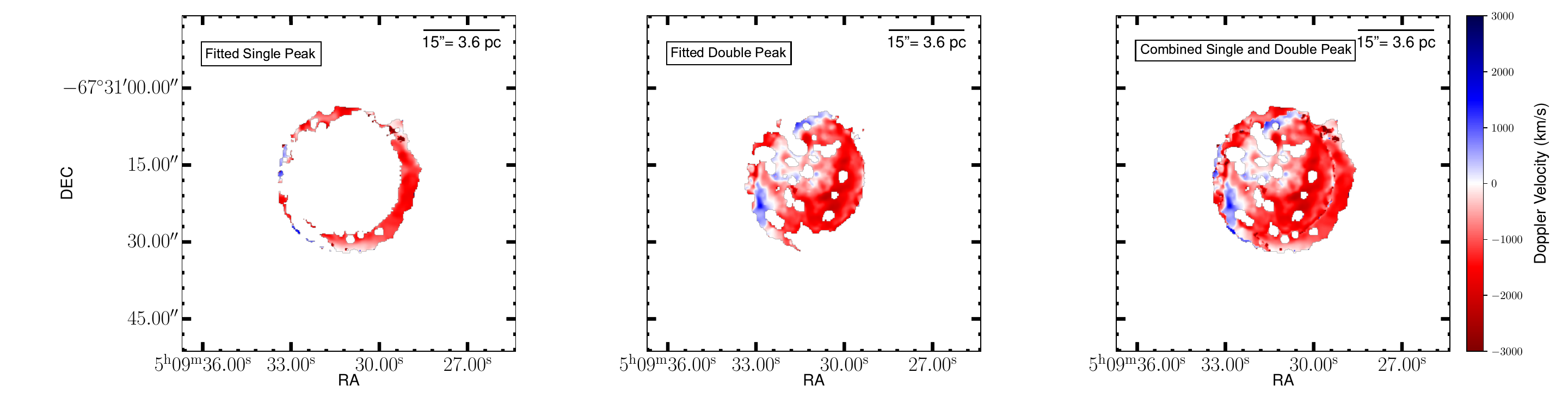}

    \caption{{\bf Left:} Doppler velocities of single-peaked profiles at the edge of the remnant, exhibiting a bulk redshift.\\
    {\bf Middle:} Doppler velocities derived from the midpoint for the case of two detected peaks, exhibiting a bulk redshift.\\
    {\bf Right:} Combined velocity map and histogram of all spaxels exhibiting [Fe\,\textsc{xiv}] ejecta, showing a median redshift of ${{\sim}}1000~\mathrm{km~s^{-1}}$. The corresponding velocity histograms for all the panels are shown in Fig.~S4}
    \label{fig:orbital_velocity}
\end{figure*}

\begin{figure*}
    \centering
    \includegraphics[width=0.6\linewidth, trim={10 10 15 0}, clip]{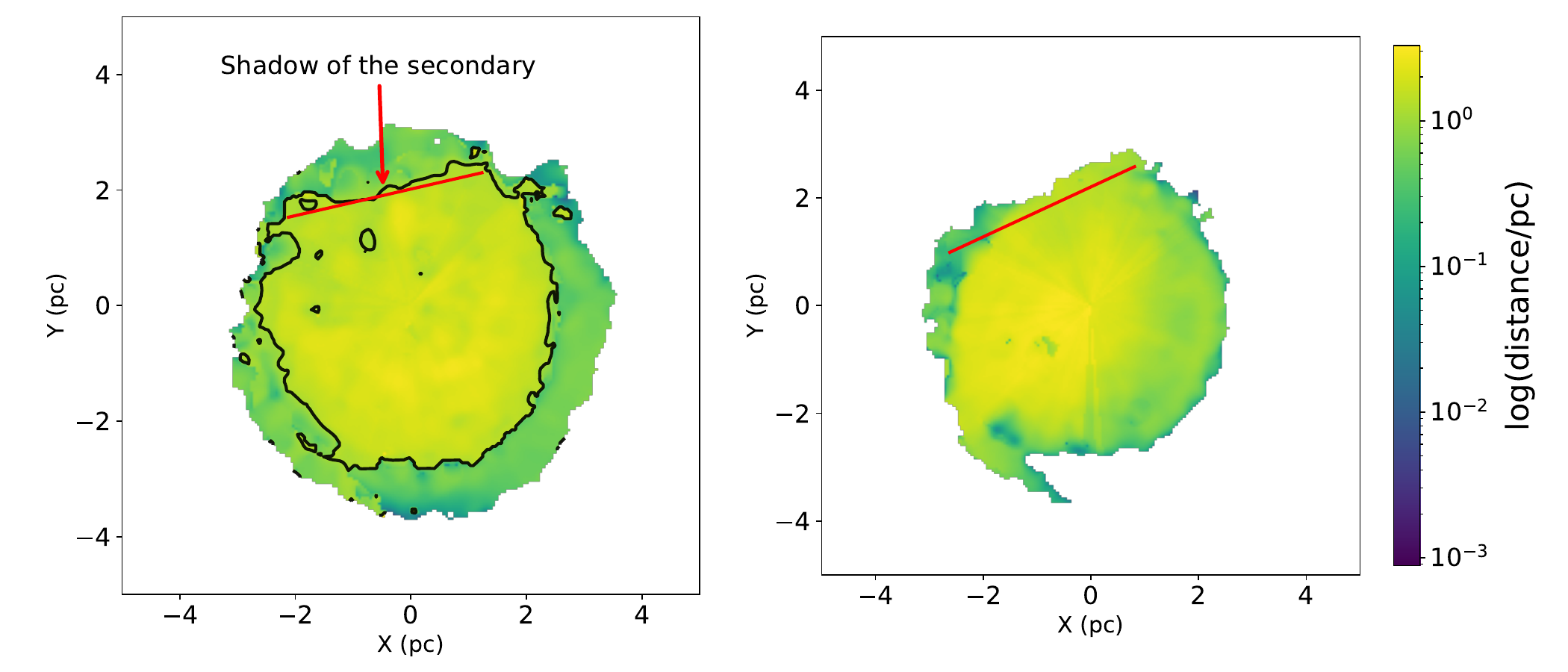}
    \includegraphics[width=0.3\linewidth, trim={230 0 0 0}, clip]{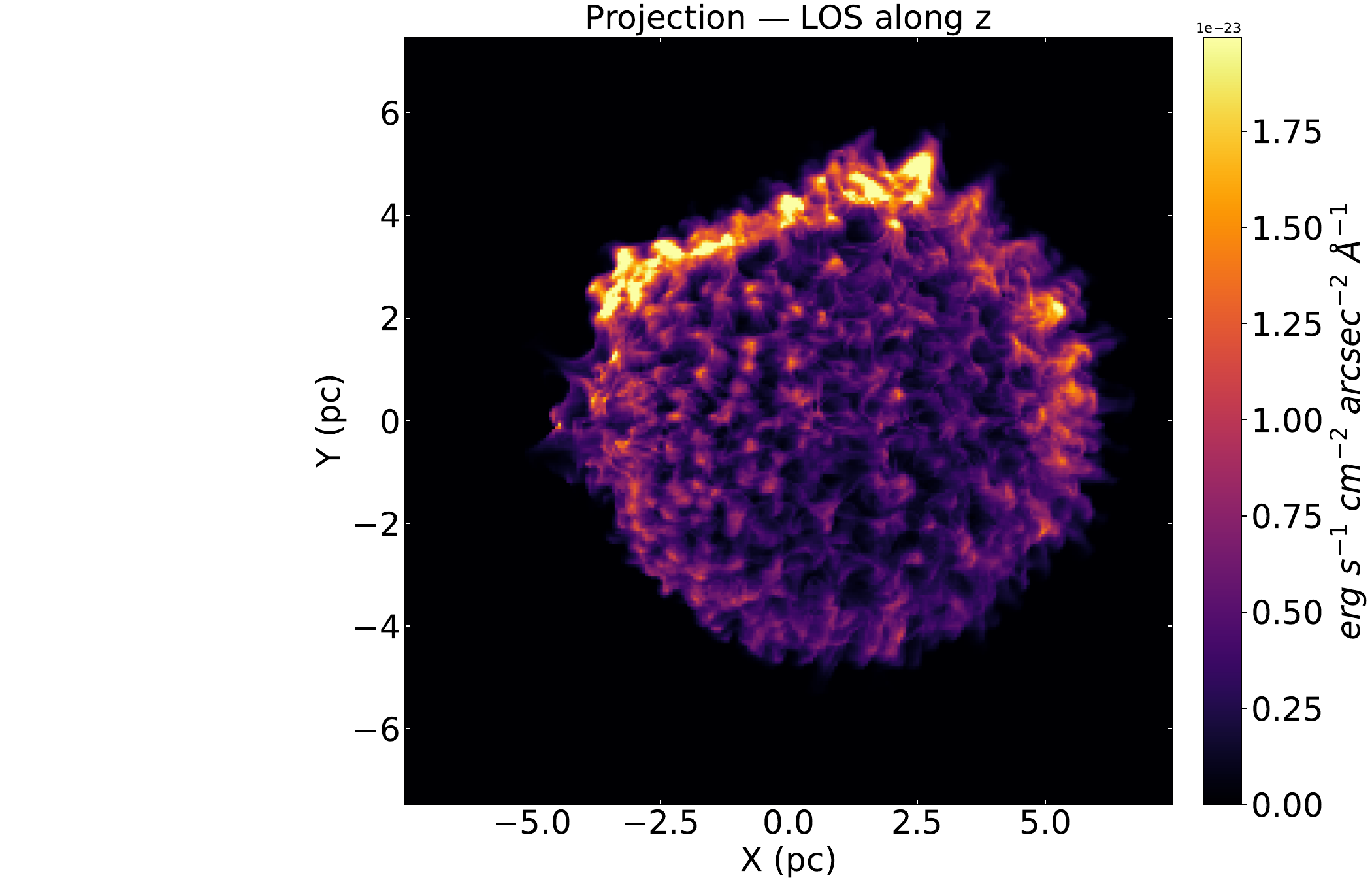}

    \caption{{\bf Left:} departure from spherical symmetry is evident in the highly red-shifted ejecta. This feature becomes more pronounced when the 1.15 pc iso-contour is overplotted to delineate the low-redshift outer edge of the ejecta. {\bf Middle:} The blue-shifted ejecta map in spatial coordinates exhibits similar deviation from spherical symmetry, with a prominent, nearly flat edge in the north-eastern region. A red line is overplotted to guide the eye and to compare the flatness against a circular edge. {\bf Right:} Projected surface brightness map of [Fe\,\textsc{xiv}] from a 3D forward modelled D6 explosion evolved to the remnant stage at an age of 500 yr, viewed along z-axis. The flattened edge associated with the “shadow of the companion” effect is visible in the north-eastern region of the remnant. }
    \label{fig:shadow}
\end{figure*}
\begin{figure}
    \centering
    \includegraphics[width=0.9\linewidth, trim={230 0 180 0}, clip]{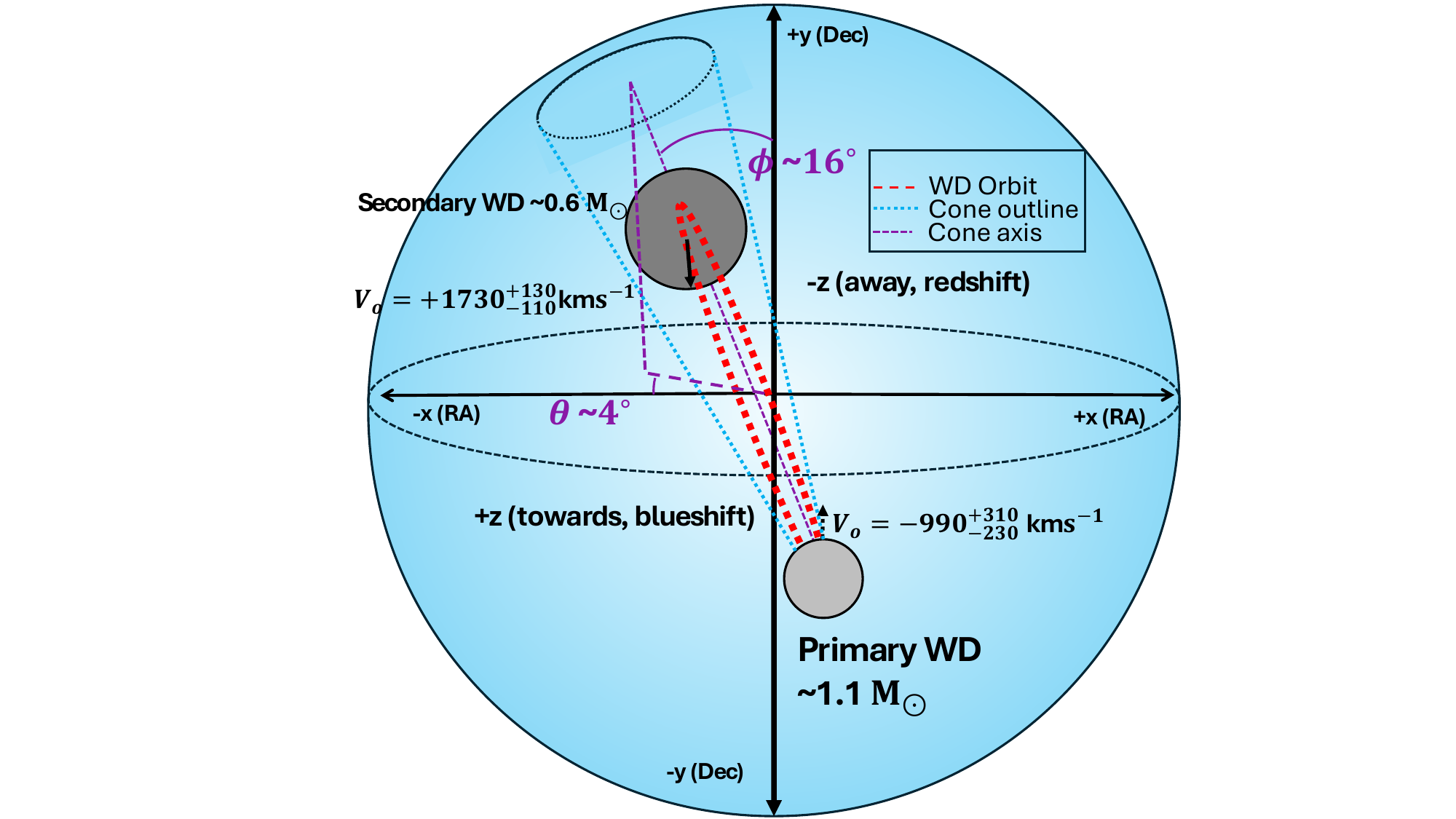}

    \caption{Three-dimensional schematic configuration of the progenitor binary system of SNR 0509-67.5 at the moment of explosion (with the observer as the reader), inferred from the cone geometry and the bulk Doppler velocity of the [Fe\,\textsc{xiv}] ejecta. The exploded primary WD (smaller light-grey sphere) and the companion WD (larger dark-grey sphere) share a binary configuration aligned along the axis of the shadow cone (dashed purple line). The secondary WD is tilted by $4^\circ$ away from the observer while moving toward us, with the primary oriented in the opposite sense. The orbital plane, shown by the red dashed line, lies nearly perpendicular to the plane of the sky, indicating that the system is viewed nearly edge-on. The line connecting the primary and secondary WDs is oriented at an angle of $16^\circ$ east of the north--south direction. }
    \label{fig:orientation}
\end{figure}

\section{Results and Discussion}
Our deep MUSE observations of SNR 0509–67.5 provide the first spatially resolved detection of Doppler-split broad coronal emission lines across an entire supernova remnant in the optical. The splitting varies systematically with projected radius, becoming increasingly prominent from the rim toward the centre.
In X-rays, high-resolution spectroscopy has previously resolved Doppler-split emission from intermediate-mass elements, enabling the construction of velocity maps and revealing the three-dimensional kinematics of young supernova remnants \citep{Willingale2002, Uchida2024, Godinaud2025}.
In the optical, spectroscopy of oxygen-rich supernova remnants such as Cassiopeia A, 1E 0102.2-7219, N132D, and SNR 0540-69.3 has revealed complex, and in some cases multi-component, velocity structure in ejecta knots and filaments \citep[e.g.][]{Vogt2011, Milisavljevic2013, Vogt2017}.
However, these observations are generally limited to localized structures and to lower-ionisation species.
In contrast, our data reveal spatially resolved double-peaked coronal line profiles throughout the reverse-shocked ejecta, with signal-to-noise ratios of ${\sim} 10$ in [Fe~\textsc{xiv}]. These profiles correspond to distinct red-shifted and blue-shifted components and are naturally explained by the projection of a spherically expanding shell along the line of sight, where emission from the approaching and receding sides contributes to the same spatial element.

\subsection{Bulk Doppler velocity in the SNR 0509-67.5 ejecta}
The line splitting does not appear at the edge of the remnant because ejecta material at the outer boundary is moving radially outward in the plane of the sky and thus has a negligible line-of-sight velocity component. Moreover, at the centre of the remnant, where the emission line splits into two Doppler peaks from the red- and blue-shifted ejecta, we find that the midpoint between the two peaks is substantially offset from the rest wavelength of [Fe\,\textsc{xiv}] after correcting for the systemic velocity of the LMC \citep[$\sim$$293~\mathrm{km~s^{-1}}$;][]{Marel2002}. We also observe localized velocity variations that appear as lighter and darker patches across the ejecta; most notably, small regions in the northern and eastern portions exhibit a systematic bulk blue-shift.  Combining the two maps therefore yields a bulk systematic red-shift velocity of ${{\sim}}1000~\mathrm{km~s^{-1}}$ for the remnant (see Fig.~\ref{fig:orbital_velocity} and Fig. S4). Multi-dimensional hydrodynamical explosion models with off-centre detonations exhibit initial asymmetries \citep{Roepke2012,seitenzahl2013a,Tanikawa2019,shen2021a}; however, when these models are evolved to the remnant phase at an age of ${\sim}$500 years, the asymmetries largely dissipate and the ejecta distribution becomes more spherical \citep{Ferrand2019,Ferrand2020}. This behaviour does not hold in the double-degenerate merger scenario. In such systems, the progenitor consists of two white dwarfs in a merging orbit, with orbital velocities reaching $1000$–$2000~\mathrm{km~s^{-1}}$. When the primary white dwarf detonates, this velocity is imparted to the inner ejecta layers, producing a systematic bulk motion \citep{pakmor2022a}. This imprint can persist into the ${\sim}$500-year remnant phase, manifesting as asymmetric expansion in forward models \citep[see][]{Ferrand2021,Ferrand2022a}. Therefore, the similar bulk Doppler shift observed in the ejecta of SNR 0509 strongly points toward a double-degenerate origin and traces the line-of-sight component of the orbital velocity of the primary WD to $-1000\pm60\mathrm{km~s^{-1}}$. The bulk velocity is defined as the median of the combined single- and double-peak velocity distribution. The quoted uncertainty includes both the asymptotic standard error of the distribution, $1.2533\sigma/\sqrt{N}$, where $\sigma$ is the velocity dispersion and $N$ is the number of spaxels in the combined map, and a systematic component ($\sigma_{\rm sys}$) derived from the dispersion in the median velocities measured from the three independent velocity maps.

\subsection{Shadow of the companion}
Type Ia supernova remnant models of the dynamically driven double degenerate double detonation (D6) and quadruple-detonation scenarios \citep{Ferrand2022a, Ferrand2025} predict the presence of signatures from the interaction of supernova ejecta with the companion. In both cases, regardless of whether the companion survives, it casts a conical shadow on the expanding ejecta. The interaction of the companion star with the highest-velocity expanding ejecta creates a conical cavity (an underdensity of ejecta material). Depending on the viewing orientation, this structure can manifest as an asymmetric or flattened edge in the projected morphology of an otherwise spherical remnant (see Fig~\ref{fig:shadow}, right panel). This effect has also been demonstrated in recent 3D hydrodynamic supernova remnant simulations by \citet{Prust2026}, who investigate the evolution of ejecta structure and composition under different scenarios: no companion, a companion that also explodes, or a surviving companion. These investigations demonstrate that departures from spherical symmetry persist throughout the remnant phase and remain pronounced even at later evolutionary stages. The interaction of the companion with the expanding ejecta leaves characteristic asymmetries in the ejecta morphology across these simulations, although the exact projected appearance in observations depends on the viewing orientation of the remnant.

Separating the blue-shifted and red-shifted ejecta of SNR 0509 may reveal this long speculated geometric signature embedded within it. The blue-shifted maps reveal a flat edge on the north-eastern part, exhibiting a deviation from its usual spherical symmetry. A similar flat edge and deviation from spherical symmetry is evident at the north eastern part of the high velocity red-shifted ejecta (see Fig.~\ref{fig:shadow}). The red-shifted ejecta map shows an apparent ring of low-velocity emission (${{\sim}}1000~\mathrm{km~s^{-1}}$) extending ahead of the sharp boundary of the high-velocity core (Fig.~\ref{fig:doppler_maps}). This does not correspond to a real discontinuity in Doppler velocity between the core and the outer ejecta. In this outer region, the [Fe\,\textsc{xiv}] line profile is not resolved into two kinematic components and is therefore fit with a single Gaussian, whose centroid simply tracks the midpoint of the unresolved red- and blue-shifted components, rather than a distinct kinematic component. Because we adopted the [Fe\,\textsc{xiv}] rest wavelength in the LMC frame without correcting for this bulk motion, the apparent emission ahead of the ejecta marked by the 1.15 pc iso-contour is a direct consequence of the systemic Doppler shift of the SNR. 

We constrain the three-dimensional orientation of the shadow cone via Bayesian inference. We selected tracer points along the north-eastern edge of the red- and blue-shifted [Fe\,\textsc{xiv}] ejecta maps, where the flat edge is most prominent and performed a 3D fit to a circular cone under the constraint that all points lie on its lateral surface and the apex is fixed at the dynamical centre mentioned in \citet{arunachalam2022a}. The cone is described by its half-opening angle $\alpha$ and two angles ($\theta$, $\phi$) defining the orientation of its symmetry axis: $\theta$ is the polar angle (tilt of the cone) measured from the observer's line of sight and $\phi$ is the azimuthal angle in the plane of the sky, measured north of west. We adopt a Gaussian likelihood where $\sigma = 0.5$~pc is a fixed positional uncertainty estimated from the spaxel scale combined with ambiguity in tracer point extraction. We report the marginalised posterior medians with $1\sigma$ error where $\theta = 86.5^{+1.9}_{-2.1}$, $\phi=105.5^{+2.1}_{-2.1}$ and $\alpha = 36.7^{+1.5}_{-1.5}$ measured in degrees (see Fig.~S5). The recovered orientation of the cone from a simple Bayesian inference lies very close to the plane of the sky with a tilt of ${\sim} 16^\circ$ towards east from the north and ${\sim} 4^\circ$ towards the red-shifted face.

\subsection{Geometry of the system}

The orientation of the cone directly translates into the 3D configuration of the progenitor binary at the moment of explosion. The azimuthal angle (${\sim} 16^\circ$ east of north) of the cone axis implies that the secondary lay to the north-northeast of the primary, with the primary's pre-explosion position correspondingly to the south-southwest of the centre of mass. The tilt (${\sim}4^\circ$ away from us) implies that both stars were nearly equidistant from us, with the primary slightly towards us but moving away and vice versa for the secondary. 
Using the opening angle derived from the cone fit and applying Eggleton’s formula \citep{Eggleton1983}, which is reasonable to assume just prior to the secondary filling its Roche lobe, we calculate a mass ratio of $q = 0.572^{+0.135}_{-0.108}$, where the mass of the primary is taken as 1.1\,$\mathrm{M_\odot}$ \citep{Mandal2026}. This corresponds to a secondary white dwarf mass of ${\sim} 0.60^{+0.15}_{-0.12}\,\mathrm{M_\odot}$, with a radius of $0.014^{+0.002}_{-0.002}~\mathrm{R_{sun}}$, an orbital separation of $0.042^{+0.008}_{-0.007}~\mathrm{R_{sun}}$ and an orbital velocity of $1730^{+130}_{-110}~\mathrm{km~s^{-1}}$ (see Table~\ref{tab:binary_parameters} for different progenitor system parameters). This corresponds to an orbital velocity of $990^{+310}_{-220}~\mathrm{km~s^{-1}}$ for the primary, which agrees within the uncertainties with the bulk Doppler redshift ($-1000\pm60~\mathrm{km~s^{-1}}$). The shadow cone constrains the binary axis but by itself is insensitive to the orientation of the orbital plane, which remains free to rotate about the cone axis. However, the fact that the observed bulk redshift is consistent, within uncertainties, with the inferred orbital velocity indicates that the orbital plane is viewed close to edge-on. Accounting for the uncertainty on the inferred orbital velocity from the cone provides an upper limit of ${\sim}40^\circ$ on the possible clockwise or anticlockwise rotation of the orbital plane from the line of sight and about the cone axis. (see Fig.~\ref{fig:orientation} for a schematic diagram of the orientation of the binaries at the moment of explosion).

\begin{table}
\centering
\caption{Derived progenitor system parameters (secondary WD masses and orbital velocity ($\mathrm{V_o}$) of the primary WD for different primary WD masses.}
\begin{tabular}{ccc}
\hline
Primary Mass ($M_\odot$) & Secondary Mass ($M_\odot$) & Primary $\mathrm{V_o}$ (km\,s$^{-1}$) \\
\hline
0.9 & $0.51^{+0.12}_{-0.10}$ & $870^{+260}_{-200}$ \\
\\
1.0 & $0.57^{+0.13}_{-0.11}$ & $950^{+290}_{-220}$ \\
\\
1.1 & $0.60^{0.15}_{-0.12}$ & $990^{+310}_{-220}$ \\
\hline
\end{tabular}
\label{tab:binary_parameters}
\end{table}
\section{Conclusion}
Deep MUSE observations of the young supernova remnant SNR 0509 reveal the splitting of the broad coronal emission line of [Fe\,\textsc{xiv}] from the reverse-shocked ejecta. This enables reconstruction of the three-dimensional ejecta distribution across the remnant. Type Ia supernova remnants typically expand into a low-density ambient medium, preserving imprints of their explosion mechanism and progenitor system. We identify imprints like the bulk Doppler shift and the flat edge on the north-eastern region of the SNR 0509, which provides new constraints on its progenitor system. 

We also report the detection of the predicted geometric signature of the companion imprinted on the [Fe\,\textsc{xiv}] ejecta, often referred to as the “shadow of the companion” \citep{Gray2016,Tanikawa2018,Tanikawa2019}. Both the red- and blue-shifted faces of [Fe\,\textsc{xiv}] exhibit a deviation from circular symmetry, with a flattened edge toward the north-eastern side. Recent 3D hydrodynamical simulations of supernova remnants predict that such geometric asymmetries (flat edge on a spherically-expanding SNR) can persist for several hundred years, including at ages of ${\sim}$500 years \citep{Ferrand2022a,Ferrand2025,Prust2026}. This provides direct evidence that a close degenerate companion was present at the time of detonation of the primary white dwarf in SNR 0509. Consequently, our results disfavour explosion scenarios that do not predict such a companion at the time of explosion for this object, including the core-degenerate lonely white dwarf scenario~\citep{Livio2003,Soker2025b}, the double-degenerate merger-to-explosion delay (DD-MED) scenario~\citep{Soker2025c}, the head-on collision of two white dwarfs~\citep{Kushnir2013} and several other proposed models~\citep{Antoniadis2020,Horowitz2021,Leung2025,Leung2026}. Whether this conclusion extends to the broader population of normal Type Ia supernovae remains to be established through studies of additional remnants. Reconstructing the orientation of the cone formed by the presence of a binary companion at the time of explosion has, for the first time, enabled a complete recovery of the physical properties and three-dimensional geometry of the progenitor system at the moment of the supernova. The orbital velocity, independently inferred using two complementary approaches -- the bulk Doppler redshift coupled with the orientation and the cone geometry -- yields consistent results that agree within their respective uncertainties.\\
Finally, our analysis concludes that SNR 0509 is consistent with originating from a double WD merger scenario, where the WDs were in a close orbit before the primary exploded. Such double-degenerate systems are increasingly emerging as some of the strongest candidates for the progenitors of normal Type Ia supernovae \citep{Polin2026}. Combined with the absence of any detected high-velocity surviving WD companion in SNR 0509 \citep{Schaefer2012,Pan14,Shields2023}, it is plausible that the secondary WD for this system also exploded. 
 Considering that any surviving companion that may exist has thus far eluded detection, this interpretation is especially justifiable because a surviving companion in a D6 system is predicted to remain highly-luminous ($\sim$$10^3$--$10^4~\mathrm{L_\odot}$) for an extended period after the explosion \citep{Bhat2025}. Moreover, recent 3D hydrodynamical simulations that model the explosion of a secondary white dwarf within the ejecta of the primary predict only minor differences in the early-time spectra, while producing a more slowly declining light curve \citep[see][for a detailed comparison]{pakmor2022a}. The overall ejecta geometry remains largely unchanged, with the main difference that the inner ejecta becomes enriched in intermediate-mass elements when the secondary white dwarf also explodes \citep{Ferrand2025, Prust2026}. This makes it difficult to conclude the fate of the secondary by direct observations in the photospheric or the remnant phase. \\
Forward modelling of a broader range of merger scenarios evolved to the remnant phase, combined with Bayesian inference against the observations, will allow the inferred progenitor parameters and explosion geometry to be constrained with greater accuracy and precision. At present, the limited availability of such models with an explicit treatment of the companion restricts our analysis to a geometric (cone-based) inference framework. In addition, the fate of the secondary white dwarf still remains uncertain in our analysis and requires further investigation, potentially through future observations with space-based spectrographs such as NIRSpec, MIRI or XRISM. Additionally, deeper integral-field spectroscopic observations of Type Ia SNRs, such as Kepler, Tycho, SNR 0519–69.0, etc are encouraged to improve the statistical understanding of how SNe Ia explode. Studies of the ejecta distribution in young supernova remnants therefore hold the key to understanding the origin and explosion physics of Type Ia supernovae as a whole.

\section*{Acknowledgments}
P.D. and I.R.S. thank Roland Crocker for useful discussions.
P.D acknowledges the use of \texttt{astropy} \citep{astropy}, \texttt{tensorflow} \citep{tensorflow2015-whitepaper} and \texttt{matploblib} \citep{hunter07} for the research analysis. The work of F.K.R.\ is supported by the Klaus Tschira Foundation and 
by the Deutsche Forschungsgemeinschaft (DFG, German Research
Foundation) -- RO 3676/7-1, project number 537700965.
F.K.R.\ acknowledges funding by the European Union (ERC, ExCEED, project number 101096243). Views and opinions expressed are however those of the authors only and do not necessarily reflect those of the European Union or the European Research Council Executive Agency. Neither the European Union nor the granting authority can be held responsible for them.

\section*{Data Availability}
This work is based on archival observations obtained with MUSE on the VLT under P.IDs 096.D-0352(A) and 0104.D-0104(A). The raw data is publicly available from the ESO Science Archive \href{archive}{https://archive.eso.org/cms.html}. Additional reduced data and analysis scripts used in this study are available from the corresponding author upon request. The machine learning code is available at \href{github}{https://github.com/Pdas888/muse-peak-classifier}.

\bibliographystyle{mnras}
\bibliography{mainbib,snr}

\setcounter{figure}{0}
\renewcommand{\thefigure}{S\arabic{figure}}
\section*{Supplementary Materials}
\FloatBarrier

\begin{figure*}
    \centering
    \includegraphics[width=0.7\linewidth, trim={0 0 0 0}, clip]{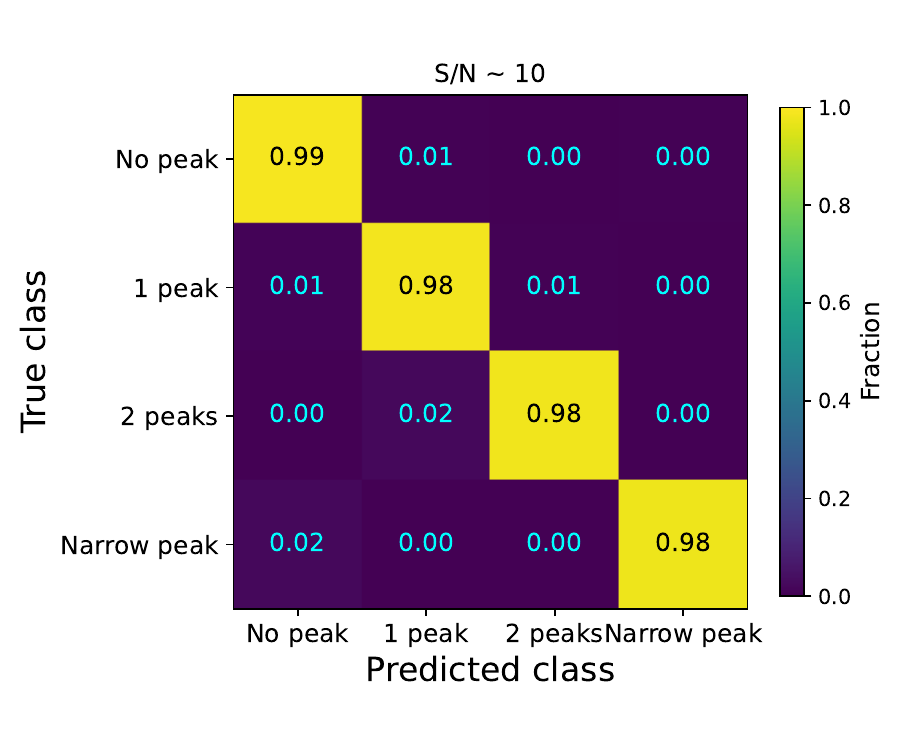}

    \caption{Confusion matrix of the interconnected neural network used for classifying and predicting peak positions. The signal-to-noise ratio (S/N) used in training the network is ${{\sim}}10$, motivated by the observed S/N of [Fe\,\textsc{xiv}] in SNR 0509–67.5. The network correctly classifies 0.99 of all no-peak cases, 0.98 of single-peak cases, 0.98 of double-peak cases, and 0.98 of narrow-peak cases in the validation dataset. Misclassifications include 0.01 of no-peak cases being identified as single-peak, 0.01 of single-peak cases being classified as either no-peak or double-peak, 0.02 of double-peak cases being identified as single-peak, and 0.02 of narrow-peak cases being classified as no-peak.}
    \label{fig:confusion}
\end{figure*}

\begin{figure*}
    \centering
    \includegraphics[width=0.8\linewidth, trim={0 0 0 0}, clip]{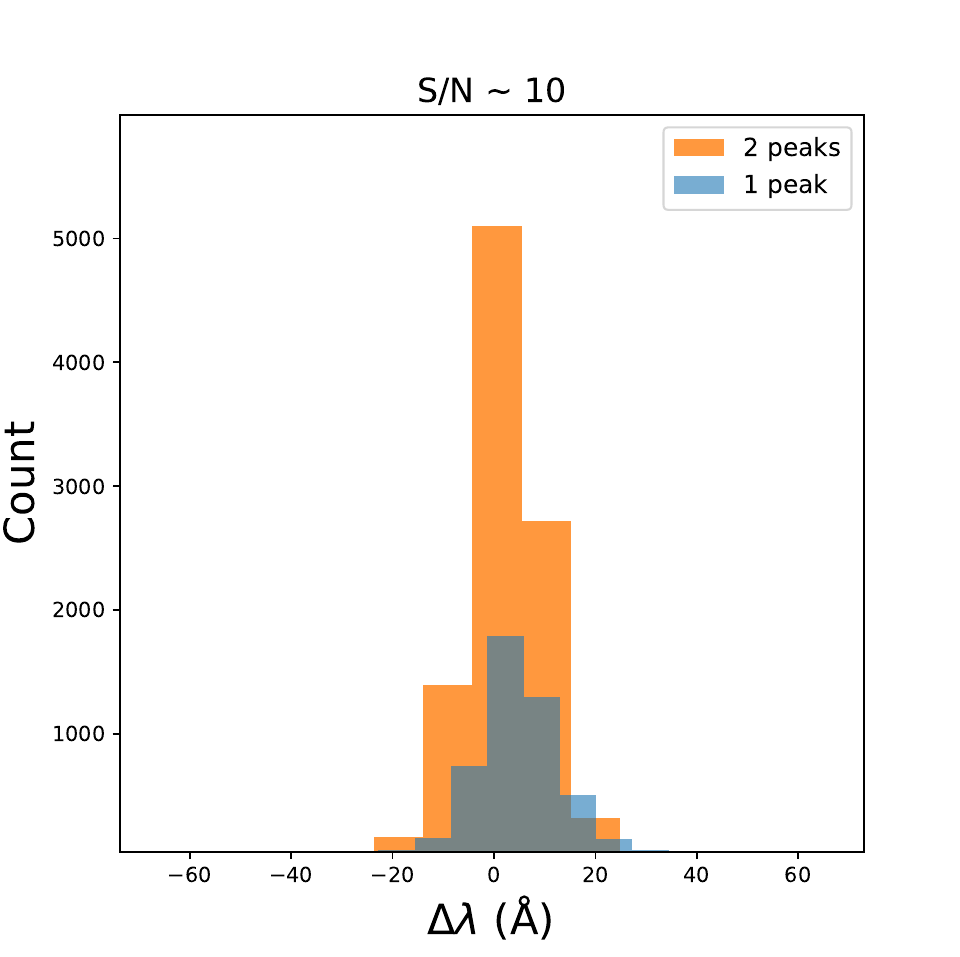}

    \caption{Histogram distribution showing the root-mean-square error (RMSE) in identifying peak locations for true single-peak and double-peak cases. Approximately 6000 spectral positions were accurately identified for the double-peak class, with a smaller number of peak positions identified with an error distributed between $-30$\AA\ and $30$\AA. For the single-peak class, approximately ${\sim}$1700 spectra were accurately identified, while ${\sim}$1600 spectra show an error of within 5\AA. The error distribution for the single-peak cases spans from $-30$\AA\ to $20$\AA.}
    \label{fig:rmse}
\end{figure*}

\begin{figure*}
    \centering
    \includegraphics[width=0.8\linewidth, trim={400 50 0 0}, clip]{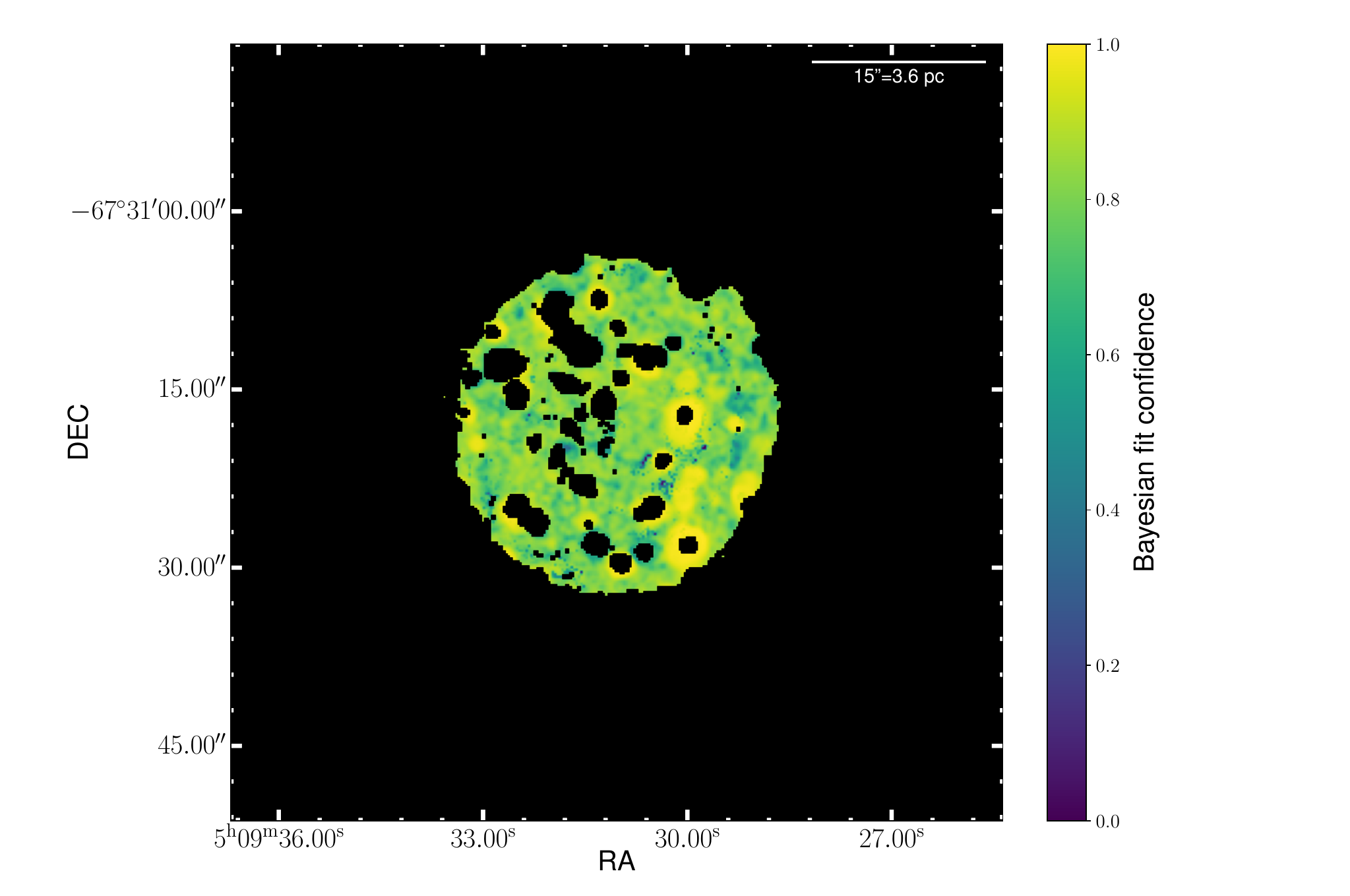}

    \caption{Map of [Fe\,\textsc{xiv}] pixels from the reverse-shocked ejecta of SNR 0509–67.5, where at least a single broad emission peak is detected in the spectra and a curve fit is performed using Bayesian inference. The fitting procedure is applied on a per-pixel basis to recover the peak parameters across the spatial extent of the ejecta. The map shows a normalised confidence score for the Bayesian fit and is displayed for the corresponding pixels, providing a measure of the reliability of the fit in each region. The map shows an average confidence score of ${{\sim}}0.8$ across the detected emission regions.}
    \label{fig:score}
\end{figure*}

\begin{figure*}
    \centering
    \includegraphics[width=1\linewidth, trim={50 10 0 0}, clip]{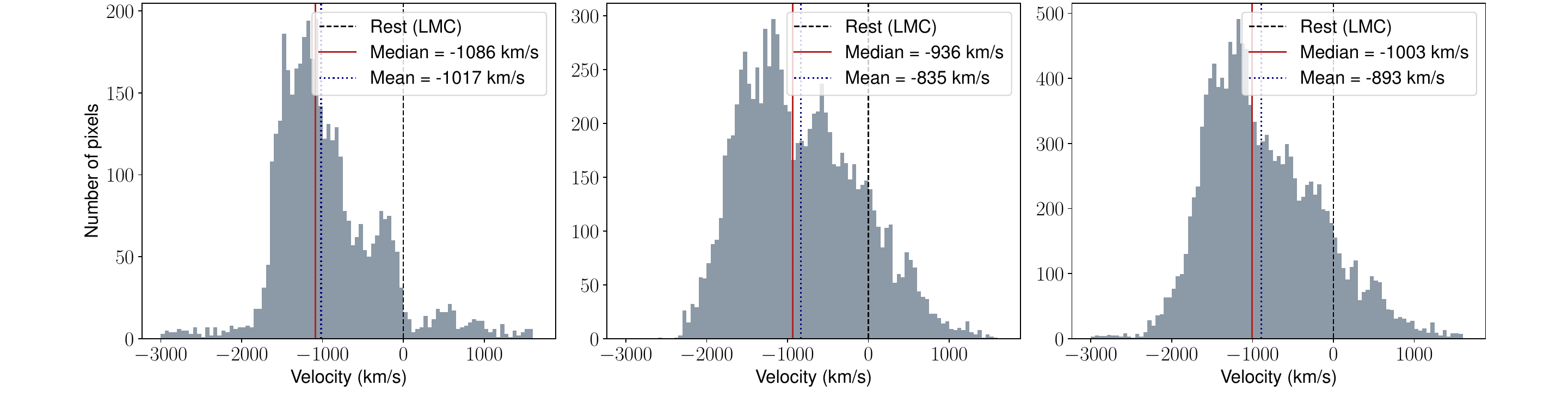}

    \caption{{\bf Left:} Histogram of Doppler velocities for all corresponding spaxels, with a median velocity of ${{\sim}}1080~\mathrm{km~s^{-1}}$.\\
    {\bf Middle:} Histogram of spaxels with double-peaked profiles, showing the Doppler shift of their midpoints, with a median velocity of ${{\sim}}940~\mathrm{km~s^{-1}}$.\\
    {\bf Right:} Combined histogram of all spaxels exhibiting [Fe\,\textsc{xiv}] ejecta, showing a median redshift of ${{\sim}}1000~\mathrm{km~s^{-1}}$.}
    \label{fig:hist}
\end{figure*}

\begin{figure*}
    \centering
    \includegraphics[width=1\linewidth, trim={0 0 0 0}, clip]{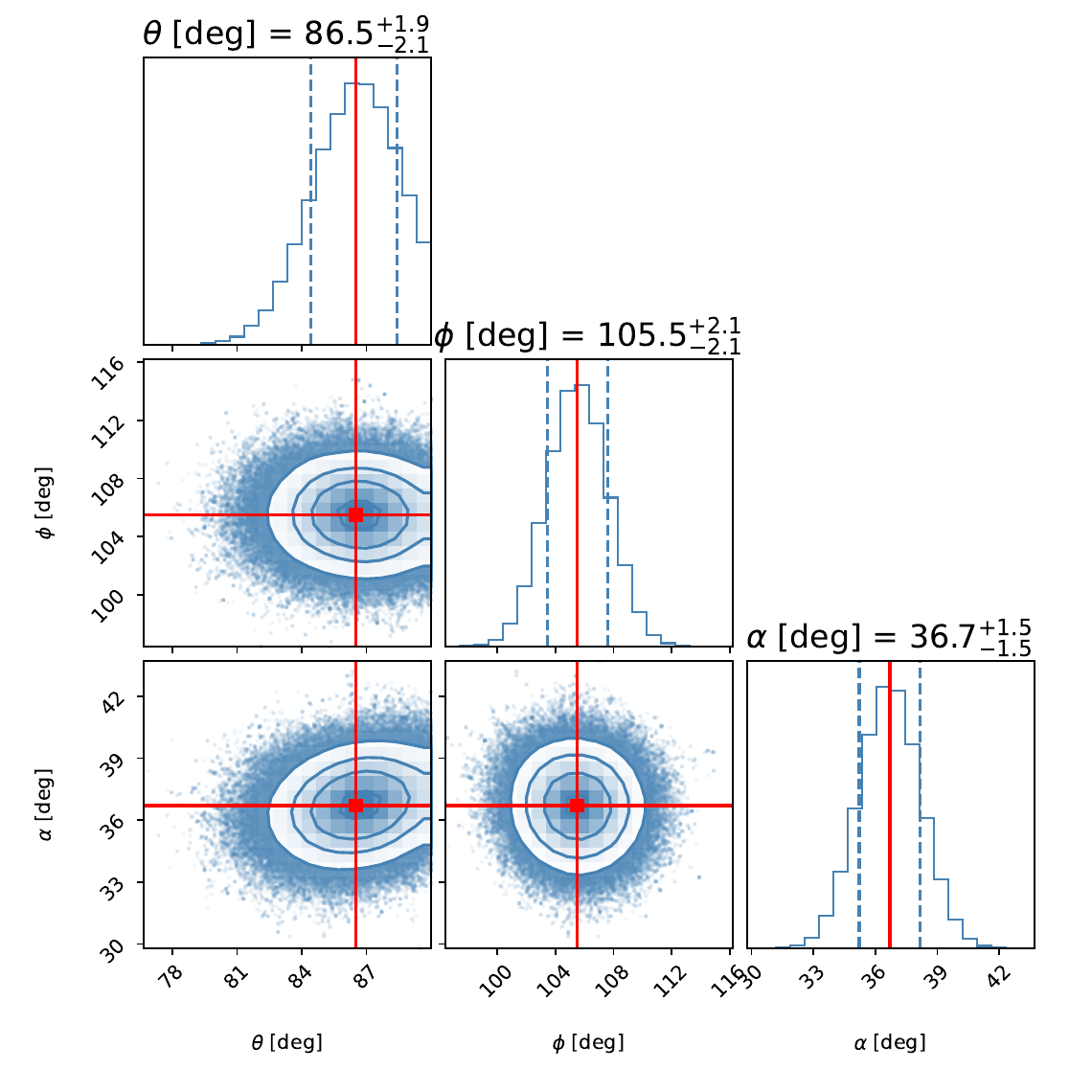}

    \caption{Corner plot from the cone fitting algorithm using Bayesian inference showing the posterior probability distribution function for the three parameters of the cone -- $\theta, \phi$ and $\alpha$. The marginalized parameter estimates are shown at the top of each panel with their $1\sigma$ confidence intervals.}
    \label{fig:corner_plot}
\end{figure*}
\end{document}